\documentclass[amsmath,amssymb,nofootinbib,prd]{revtex4}

\usepackage{setspace}
\usepackage{graphicx}

\usepackage{amsmath,amssymb,amsfonts,amsthm}
\usepackage{enumerate}

\begin{document}

\title{Loop quantum gravity of a spherically symmetric scalar field coupled to gravity with a clock}

\author{Rodolfo Gambini$^1$,  Jorge Pullin$^2$}
\affiliation{1. Instituto de F\'{\i}sica, Facultad de Ciencias, Igu\'a 4225, esq. Mataojo,
11400 Montevideo, Uruguay. \\
2. Department of Physics and Astronomy, Louisiana State University,
Baton Rouge, LA 70803-4001, USA.}

\begin{abstract}
The inclusion of matter fields in spherically symmetric loop quantum gravity has proved problematic at the level of implementing the constraint algebra including the Hamiltonian constraint. Here we consider the system with the introduction of a clock. Using the Abelianizaton technique we introduced in previous papers in the case of gravity coupled to matter, the system can be gauge fixed and rewritten in terms of a restricted set of dynamical variables that satisfy simple Poisson bracket relations.   This creates a true Hamiltonian and therefore one bypasses the issue of the constraint algebra. We show how loop quantum gravity techniques may be applied for the vacuum case and show that the Hamiltonian system reproduces previous results for the physical space of states and the observables of a Schwarzchild black hole.
\end{abstract}
\maketitle

\section{Introduction}
Spherically symmetric loop quantum gravity has proved to be an interesting symmetry reduction that allows the study of black holes, singularity elimination by the quantum theory, and other issues \cite{usreview}. However, the introduction of matter has proved problematic. In the vacuum theory a redefinition of the constraints allows to turn them into a Lie algebra and complete the Dirac quantization. At present it is not known how to carry on this procedure when one couples matter fields, even simple ones like a massless scalar field, at the quantum level. Being able to include massless scalar fields is important, as it gives rise to a rich dynamics that not only includes black hole formation but also the critical phenomena discovered by Choptuik \cite{choptuik}. 

Here we would like to consider a spherically symmetric massless scalar field coupled to spherically symmetric gravity in the presence of a clock given by a second scalar field. The latter gives rise to a true Hamiltonian and this allows to sidestep the issue of the Hamiltonian constraint by quantizing a gauge fixed theory. There is some literature on the use of matter clocks in quantum gravity (see \cite{gravityquantized} for references). Here we are exploiting the advantages of the simplifications due to spherically symmetric gravity to make progress in defining the relevant quantum operators in a precise way. 

The organization of this article is as follows: in section 2 we discuss the classical theory for spherically symmetric gravity coupled to a clock and identify the Hamiltonian. In section 3 we reproduce the calculations for gravity coupled to a scalar field in the presence of a clock. In section 4 we discuss the quantization of the pure gravity case coupled to a clock and show that one recovers, in the weak gravity approximation, previous results obtained using Dirac's procedure. We end with conclusions.

\section{Classical theory: gravity with a scalar clock}

As is usual in spherically symmetric loop quantum gravity one has as canonical variables the triads in the radial and tangential directions $E^x, E^\varphi$ and their canonical momenta $K_x,K_\varphi$. The variables for the scalar field are $\phi$ and $P_\phi$. We refer the reader to \cite{usreview,florencia}  for more details. We start by a total Hamiltonian constraint similar to the ones considered in the papers cited, involving a scalar field that we will take as a clock (we take $\hbar=c=1$),
\begin{eqnarray}
    H_T&=&\frac{1}{G}\int dx \left[ N^x\left(\left(E^x\right)'K_x-E^\varphi\left(K_\varphi\right)'-8\pi P_\phi \phi'\right)\right.\nonumber\cr
    &&+N
    \left(
    -\frac{E^\varphi}{2\sqrt{E^x}}-2\sqrt{E^x}K_\varphi K_x-\frac{K_\varphi^2 E^\varphi}{2\sqrt{E^x}}
    +\frac{\left(\left(E^x\right)'\right)^2}{8\sqrt{E^x}E^\varphi}
    -\frac{\sqrt{E^x}\left(E^x\right)'\left(E^\varphi\right)'}{2\left(E^\varphi\right)^2}\right.\nonumber\cr
    &&\left.\left.+\frac{\sqrt{E^x}\left(E^x\right)''}{2 E^\varphi} +\frac{2\pi G P_\phi^2}{\sqrt{E^x}E^\varphi}
    +\frac{2\pi G\sqrt{E^x}{E^x}\left(\phi'\right)^2}{E^\varphi}\right)\right],
\end{eqnarray}
where prime is derivative with respect to $x$, the radial coordinate.

We now proceed to rescale the lapse and the shift as we have done in previous work \cite{usreview}, which leads in vacuum to an Abelian algebra of constraints where the vacuum part of the Hamiltonian constraint takes the form
\begin{equation}
H_{\rm new} :=\frac{\left(E^x\right)'}{E^\varphi}H-2\frac{\sqrt{E^x}}{E^\varphi}K_\varphi H_x= -\frac{1}{G}\left[ \sqrt{E^x}\left(1-\frac{[(E^x)']^2 }{4 (E^\varphi)^2}+K_\varphi^2\right)\right]', \label{1}
\end{equation}
where we call $H_{\rm new}$ the Hamiltonian constraint that results after the rescaling. Notice that it is independent of $K_x$.
As it is usual in the spherically symmetric case we end up with a Hamiltonian and a diffeomorphism first class constraints. We will work with a totally gauge fixed system. We start by choosing 
 $E^x(x,t)=E^x(x)$, where $E^x(x)$ is a given function of $x$ independent of time. 
 This choice sets, after imposing the preservation of the gauge, the new shift to zero, and determines $K_x$,
\begin{equation}
    K_x=\frac{1}{\left(E^x\right)'}
    \left(
    E^\varphi K_\varphi' +8\pi G P_\phi\phi'
    \right).
\end{equation}
To see that that the shift is zero is to impose the preservation in time of the gauge choice $E^x(x,t)=E(x)$. Since $K_x$ does not appear in the abelianized Hamiltonian constraint, the Poisson bracket of the gauge choice with the diffeomorphism constraint (which has to be zero) is proportional to the shift.

The gravitational part of the Hamiltonian constraint can be written as a total derivative, that vanishes in vacuum, since it is a combination of the original constraints,
\begin{equation}
    H_{G}=N_{\rm new}\left[
    -\sqrt{E^x}\left(
    1+K_\varphi^2 -\frac{\left(\left(E^x\right)'\right)^2}{4 \left(E^\varphi\right)^2}\right)+ 2GM
    \right]'\equiv N_{\rm new}C',
\end{equation}
with $M$ the ADM mass arising as a constant of integration (see \cite{usreview0} for details). In previous works we integrated by parts, but here do not  and therefore the mass term is irrelevant. The clock part of the Hamiltonian constraint takes the form,
\begin{equation}
    H_C=
    \frac{1}{\left(E^\varphi\right)^2\sqrt{E^x}}
    \left(
    2 N_{\rm new} \pi G \left(\left(E^x\right)' \left(E^x\right)^2 \left(\phi'\right)^2-8E^x K_\varphi E^\varphi P_\phi \phi'+\left(E^x\right)' {P_\phi}^2
    \right)
    \right).
\end{equation}
For positive lapse $H_C$ is positive and $H_G$ negative. 
We now impose a second gauge fixing  $\phi(x)=t/l_0^2$. The arbitrary constant $l_0$ characterizes the range of validity of the clock, and will be determined later when one computes the expectation value of the Hamiltonian by the size of the spatial patch to be studied in order to have a causally connected region. Preservation of the gauge fixing in time,
\begin{equation}
{\dot \phi(x)}=\left\{\phi(x),\int{N_{\rm new}(z)H_C(z)dz}\right\}=\frac{1}{l_0^2},\nonumber
\end{equation}
fixes the lapse,
\begin{equation}
    N_{\rm new}=\frac{\left(E^\varphi\right)^2\sqrt{E^x}}{4 \pi \left(E^x\right)'P_\phi l_0^2},
\end{equation} 
and the gauge fixing is only possible if $P_\phi$ is non-vanishing.
With this, the Hamiltonian constraint $H_T=H_G+H_C$ becomes second class and implies,
\begin{equation}
    \frac{C'}{\left(E^x\right)'}+\frac{2\pi G  P_\phi^2}{\left(E^\varphi\right)^2 \sqrt{E^x}}=0.\label{6}
\end{equation}

Since we have fixed $\phi$, we can solve for the conjugate momentum,
\begin{equation}
    P_\phi=\left({2\pi G}\right)^{-1/2}{\sqrt{\frac{- C'}{\left(E^x\right)'}\sqrt{E^x}} E^\varphi}.\label{77}
\end{equation}
In the resulting gauge fixed theory both the diffeomorphism and the Hamiltonian constraints are second class and allow to write $K_x$ and $P_\phi$ in terms of the vacuum dynamical variables $K_\varphi$ and $E^\varphi$.

Our previous black hole solutions in vacuum \cite{usreview} are modified by terms of order $G$, as expected, and one recovers those solutions in the limit $G=0$. The time derivative of $P_\phi$ is (with respect to time $t)$,
\begin{equation}
    \dot{P}_\phi=
    \frac{8\pi\sqrt{E^x}N_{\rm new}}{E^\varphi} 
    \left[\frac{2 K_\varphi P_\phi \left(E^\varphi\right)'}{E^\varphi}
    -2  P_\phi K_\varphi'-2 \ln(N_{\rm new})' K_\varphi P_\phi
    -2  K_\varphi P_\phi'
    -\frac{ K_\varphi P_\phi \left(E^x\right)'}{\sqrt{E^x}}
    \right],\label{7}
\end{equation}
which can be considered as a subsidiary equation that may be obtained from equation (\ref{6}) and the evolution equations for the variables that appear in the pure gravity true Hamiltonian, as we shall see. 

The true Hamiltonian reproduces via the canonical equations the evolution of $K_\varphi(x)$ and $E^\varphi(y)$ obtained following the Dirac's procedure for the constrained system.
We proceed to the determination of the evolution equation for $K_\varphi(x)$. Taking into account that $\{K_\varphi(x),E^\varphi(y)\}=G\delta^3(x-y)$, and equation (\ref{1}) we have that,
\begin{equation}
    \dot{K}_\varphi=-\frac{1}{2\left(E^\varphi\right)^3 \sqrt{E^x}}
    \left[ 8 G\pi N_{\rm new} \left(E^x\right)'P_\phi^2- E^x N_{\rm new}' \left(\left(E^x\right)'\right)^2
    \right].
\end{equation}

Substituting $N_{\rm new}$ and $P_\phi$, we get,
\begin{eqnarray}
    \dot{K}_\varphi &=&-
    \frac{\sqrt{2 G\pi}}{2 \sqrt{ \left(E^x\right)'}\left(C'\right)^{3/2}l_0^2 \left(E^\varphi\right)^3\left(E^x\right)^{1/4}}
    \left[
    4 E^x \left(\left(E^x\right)'\right)^2\left(E^\varphi\right)'C'\nonumber\right.\\
&&\left.    -2E^x\left(\left(E^x\right)'\right)^2 C'' E^\varphi-2 E^x \left(E^x\right)' \left(E^x\right)''C' E^\varphi+\left(\left(E^x\right)'\right)^3C' E^\varphi
    +16\sqrt{E^x}\left(C'\right)^2 \left(E^\varphi\right)^3
    \right]\label{10}.
\end{eqnarray}
The equation for the derivative of $E^\varphi$ is obtained in the same way, we do not show it explicitly. Note that since we have solved for $\phi$ and its conjugate momentum $P_\phi$, the Dirac brackets for $K_\varphi, E^\varphi$ coincide with the Poisson brackets, when one takes the $\phi=t/l_0^2$ and equation (\ref{6}) as strong constraints.  The redefinition of the Lagrange multipliers and the abelianization of the Hamiltonian constraint allows
to determine the lapse $N_{\rm new}$  as an algebraic function of the dynamical variables and to use Poisson brackets
instead of the more involved Dirac brackets. These properties are crucial for obtaining  a strictly canonical formulation.
The following true Hamiltonian leads to  usual Hamilton equations for $K_\varphi$ and $E^\varphi$ that are identical to the ones obtained above using the Dirac method,
\begin{equation}
    H_{\rm True}=\int_{-\infty}^{\infty} 
    \frac
    { E^\varphi}{l_0^2\sqrt{2 \pi G } }\sqrt{-\frac{C' \sqrt{E^x}}{\left(E^x\right)'}}
    dx .\label{11}
\end{equation}
This expression can be recognized by considering the gravitational part of the Hamiltonian constraint and substituting the lapse $N_{\rm new}$ and the expression for $P_\phi$.

\section{Classical theory: gravity with a scalar clock and a scalar field}

We start by considering the total Hamiltonian with a scalar field $\phi$ as a clock and a gravitating massless scalar field $\psi$. Their canonical momenta are $P_\phi$ and $P_\psi$,
\begin{eqnarray}
    H_T&=&\frac{1}{G}\int dx \left[ N^x\left(\left(E^x\right)'K_x-E^\varphi\left(K_\varphi\right)'-8\pi P_\phi \phi'-8\pi P_\psi \psi'\right)\right.\nonumber\cr
    &&+N
    \left(
    -\frac{E^\varphi}{2\sqrt{E^x}}-2\sqrt{E^x}K_\varphi K_x-\frac{K_\varphi^2 E^\varphi}{2\sqrt{E^x}}
    +\frac{\left(\left(E^x\right)'\right)^2}{8\sqrt{E^x}E^\varphi}
    -\frac{\sqrt{E^x}\left(E^x\right)'\left(E^\varphi\right)'}{2\left(E^\varphi\right)^2}\right.\nonumber\cr
    &&\left.\left.+\frac{\sqrt{E^x}\left(E^x\right)''}{2 E^\varphi} +\frac{2\pi G P_\phi^2}{\sqrt{E^x}E^\varphi}
    +\frac{2\pi G\sqrt{E^x}{E^x}\left(\phi'\right)^2}{E^\varphi}
    +\frac{2\pi G P_\psi^2}{\sqrt{E^x}E^\varphi}
    +\frac{2\pi G\sqrt{E^x}{E^x}\left(\psi'\right)^2}{E^\varphi}
    \right)\right].
\end{eqnarray}
Redefining the shift as before,
\begin{equation}
    N^x=N^r+\frac{2 N \sqrt{E^x}K_\varphi}{\left(E^x\right)'},
\end{equation}
 we get, 
\begin{eqnarray}
    H_T&=&\frac{1}{G}\int dx\left[ N
    \left(
    -\frac{E^\varphi}{2\sqrt{E^x}}
    -\frac{K_\varphi^2 E^\varphi}{2\sqrt{E^x}}
    +\frac{\left(\left(E^x\right)'\right)^2}{8\sqrt{E^x}E^\varphi}
    -\frac{\sqrt{E^x}\left(E^x\right)'\left(E^\varphi\right)'}{2\left(E^\varphi\right)^2}\right.\right.\nonumber\cr
    &&+\frac{\sqrt{E^x}\left(E^x\right)''}{2 E^\varphi} +\frac{2\pi G\phi P_\phi^2}{\sqrt{E^x}E^\varphi}
    +\frac{2\pi G\sqrt{E^x}{E^x}\left(\phi'\right)^2}{E^\varphi}
    +\frac{2\pi G\psi P_\psi^2}{\sqrt{E^x}E^\varphi}
    +\frac{2\pi G\sqrt{E^x}{E^x}\left(\psi'\right)^2}{E^\varphi}\nonumber\\
    &&\left.\left.-\frac{2\sqrt{E^x}K_\varphi} {\left(E^x\right)'}
    \left(E^\varphi K_\varphi'+8\pi G P_\phi \phi'+
    8 \pi G P_\psi \psi'
    \right)\right)
    + N^x \left(\left(E^x\right)'K_x -E^\varphi K_\varphi'
    -8\pi G \left(P_\phi\phi'+P_\psi\psi'\right)
    \right)\right],
\end{eqnarray}
and setting as before that $E^x$ is a time independent function, and therefore $N^x=0$ we have that,
\begin{equation}
    K_x=\frac{1}{\left(E^x\right)'}\left(E^\varphi K_\varphi' +8\pi G\left(P_\phi \phi'+P_\psi \psi'\right)
    \right).
\end{equation}

The gravitational part can be written (as in equation (3)) in terms of the  derivative of $C$, given by,
\begin{equation}
    C=-\sqrt{E^x}\left(1+K_\varphi^2 -\frac{\left(\left(E^x\right)'\right)^2}{4\left(E^\varphi\right)2}\right)+2 G M,
\end{equation}
and, having introduced the integration constant associated with the ADM mass $M$,  $C$ (and its derivative) vanish in vacuum. 

The Hamiltonian constraint with a scalar field and a scalar field clock, when written in terms of the re-scaled lapse $N_{\rm new}$ takes the form,

\begin{eqnarray}
    H_{\rm total} &=& \frac{1}{\sqrt{E^x}\left(E^\varphi\right)^2}
    \left(
    2 N_{\rm new} \pi\left(
    \left(E^x\right)^2\left( \left(\phi'\right)^2+\left(\psi'\right)^2\right)
    +P_\phi^2+P_\psi^2
    \right)\left(E^x\right)'
    -8E^x E^\varphi K_\varphi
    \left(\phi'P_\phi+\psi'P_\psi\right)\right)\nonumber\\
    && +\frac{N_{\rm new}}{G}\left( 
    -\frac{\left(1+K_\varphi^2
    -\frac{\left(\left(E^x\right)'\right)^2}{4 \left(E^\varphi\right)^2}\right)\left(E^x\right)'}{2\sqrt{E^x}}
    -\sqrt{E^x}
    \left(
    2 K_\varphi K_\varphi'
    -\frac{\left(E^x\right)'\left(E^x\right)''}{2 \left(E^\varphi\right)^2}
    +\frac{\left(\left(E^x\right)'\right)^2 \left(E^\varphi\right)'}{2\left(E^\varphi\right)^3}
    \right)\right),
\end{eqnarray}
which can be rewritten as, 
\begin{eqnarray}
H_{\rm total} &=&
\frac{2 N_{\rm new} \pi}{\sqrt{E^x}\left(E^\varphi\right)^2}
\left[\left(E^x\right)'\left(
\left(E^x\right)^2\left(
 \left(\phi'\right)^2+
 \left(\psi'\right)^2
\right)
+P_\phi^2+P_\psi^2
\right)
-8E^x E^\varphi K_\varphi 
\left(\phi'P_\phi+\psi' P_\psi\right)
\right]+\frac{N_{\rm new}}{G} C'.\label{17}
\end{eqnarray}

As in the previous section, we now fix the gauge, which is equivalent to choosing the clock $\phi=t/l_0^2$ and impose that it is preserved upon evolution. This determines the lapse,
\begin{equation}
    N_{\rm new}=\frac{\sqrt{E^x} \left(E^\varphi\right)^2}{4 \pi P_\phi \left(E^x\right)' l_0^2}.\label{18}
\end{equation}

The Hamiltonian constraint and the gauge fixing of the clock field become now second class,
\begin{eqnarray}
H_{\rm total} &=&
\frac{2 N_{\rm new} \pi}{\sqrt{E^x}\left(E^\varphi\right)^2}
\left[\left(E^x\right)'\left(
\left(E^x\right)^2
 \left(\psi'\right)^2
+P_\psi^2+P_\phi^2
\right)
-8E^x E^\varphi K_\varphi 
\psi'P_\psi
\right] +\frac{N_{\rm new}}{G} C',\label{19}
\end{eqnarray}
and allow us to impose strongly the constraint in the sense of Dirac and to obtain an equation for $P_\phi$,
\begin{equation}
    P_\phi=\frac{1}{\sqrt{\pi G}\left(E^x\right)'}
    \sqrt{\left[
    -\left(
    \frac{C'\sqrt{E^x}\left(E^\varphi\right)^2}{2}
    + \pi G\left(E^x\right)'
    \left[\left(E^x\right)'\left(
\left(E^x\right)^2
 \left(\psi'\right)^2
+P_\psi^2
\right)
-8E^x E^\varphi K_\varphi 
\psi'P_\psi
\right]
    \right)
    \right]} \label{20}.
\end{equation}

Having solved the second class constraints for $\phi$ and $P_\phi$, the Dirac Brackets for $\psi,P_\psi,K_\varphi,E^\varphi$ coincide with the Poisson Brackets.

We now proceed to compute the time derivatives of  $\psi,P_\psi,K_\varphi,E^\varphi$ by following the Dirac procedure, that is, by evaluating the Poisson bracket with the Hamiltonian constraint (\ref{17}) and substituting $N_{\rm new}$  and $P_\phi$ using equation (\ref{18}) and (\ref{20}). We here compute $\dot K_\varphi$, the others are similar,

\begin{equation}
    \dot{K}_\varphi = 
    -\frac{1}{2\sqrt{E^x}\left(E^\varphi\right)^3}
    \left(
    -\left(\left(E^x\right)'\right)^2 {N_{\rm new}}' E^x
    +8\pi G N_{\rm new} \left(\left(E^x\right)^2\left(\psi'\right)^2+P_\psi^2+P_\phi^2\right)\left(E^x\right)'-32\pi G N_{\rm new} E^x E^\varphi K_\varphi
    P_\psi \psi'
    \right).
\end{equation}
After the substitution of (\ref{18}) and (\ref{20}) we get
an equation that is reproduced by one of the canonical equations resulting from the following true Hamiltonian,
\begin{equation}
    H_{\rm true}=\int_{-\infty}^\infty dx
     \frac{ \sqrt{
    -C' \sqrt{E^x}\left(E^\varphi\right)^2
    -2\pi G
    \left(
    \left(E^x\right)'\left(E^x\right)^2\left(\psi'\right)^2
    -8E^x K_\varphi E^\varphi P_\psi \psi'+\left(E^x\right)'P_\psi^2
    \right)}}
    {l_0^2 \sqrt{2\pi G \left(E^x\right)'}}
\end{equation}

And it can be seen that the equations of motion for $K_\varphi, E^\varphi, \psi$ and $P_\psi$ obtained from $H_{\rm total}$ can be reproduced by $H_{\rm true}$. The computation is straightforward but laborious and does not add to the understanding of the procedure. As in the vacuum case, the Hamiltonian  can be recognized by considering the gravitational part of the Hamiltonian constraint and substituting the lapse $N_{\rm new}$ and the expression for $P_\phi$.

The presence of the square root imposes restrictions. When one
has a scalar field $\psi$ present, one has to have that,
\begin{equation}
    C'\leq 
    -\frac{2 G \pi
    \left[\left(E^x\right)'\left(
\left(E^x\right)^2
 \left(\psi'\right)^2
+P_\psi^2
\right)
-8E^x E^\varphi K_\varphi 
\phi'P_\psi
\right]}{\left(E^\varphi\right)^2\sqrt{E^x}}.\label{24}
\end{equation}
One can only use as a clock the field in configurations that satisfy this inequality. One simply chooses initial data satisfying it ,and that suffices because the Hamiltonian constraint \ref{19} ensures that the inequality is preserved. We will discuss later how to handle this in the quantum theory.

\section{Quantization of the pure gravity case: preliminaries}

It is of interest to see if the solutions of the constrained system we studied in previous papers \cite{usreview} are well approximated by the low energy states of the Hamiltonian system, in particular that we recover the vacuum solutions in an approximate fashion. Notice that equation (\ref{6}) shows that the clock modifies the Hamiltonian constraint with a term proportional to $GP_\phi^2$. Only when this term is small one recovers that vacuum constraint.

As we mentioned, the Hamiltonian involves a square root, and therefore the evolution is not defined in the complete phase space. Let us discuss this in some detail. We start by rewriting equation (\ref{6}),
\begin{equation}
C'+\frac{2\pi G  {\left(E^x\right)'} P_\phi^2}{\left(E^\varphi\right)^2 \sqrt{E^x}}=0,\label{6mod}
\end{equation}
and recalling that 
\begin{equation}
    H_{\rm true}=\int_{-\infty}^{\infty} dx {{\frac{E^\varphi}{\sqrt{2\pi G}l_0^2}} \sqrt{-\sqrt{E^x} \frac{{C'}}{(E^x)'}}},
    \label{25}
\end{equation}
we have that  $H_{\rm true}>0$ for the configurations satisfying (\ref{6mod}). We have included in the square root $\left(E^x\right)'$ that can be chosen bigger than zero for a classical black hole. Notice that when $C'>0$ the $H_{\rm true}$ and the momentum of the clock become imaginary and therefore the evolution of the system is not defined in the complete phase space. At the classical level it is enough to recall that one needs to start from initial conditions that lead to a real Hamiltonian, given the preservation of the constraints. However in order to ensure that the quantum Hamiltonian is self-adjoint we need to include the remaining region of the phase space without modifying the classical solutions of the real sector. This is ensured by considering

\begin{equation}
    H_{\rm true}=\int_{-\infty}^{\infty} dx \frac{E^\varphi}{\sqrt{2\pi G}l_0^2}
    \sqrt{\sqrt{E^x} \left\vert\frac{C'}{(E^x)'}\right\vert}.
\end{equation}
The inclusion of the modulus does not change the dynamics classically for the allowed initial values defined by the inequality. It however provides a classical theory that can be readily quantized without introducing non-self adjoint operators.  

To begin with, let us note that the solutions to the evolution equations generated by the true Hamiltonian, that approximately satisfy the vacuum constraint $C'=O({G}/{{l_0^2}})$ initially, also vanish approximately for all time since ${\dot C}' = \{C',H_{\rm True}\}=O({\sqrt{G}}/{{l_0}})$ and the Hamiltonian is (approximately) a constant of motion. Therefore the classical solutions for the
gauge $E^x(x,t)=E^x(x)$ are recovered in the limit where the effect of the clock is negligible. However, it is clear that new solutions appear in which the evolution of the gravitational variables is affected by the clock.  The original theory, with constraint $C'=0$ had an additional gauge freedom that allowed, for instance, to choose $K_\varphi(x,t)$, which corresponds to a choice of foliation. This freedom is lost because now; given $K_\varphi(x,0)$,
its evolution is determined by the choice of the clock. For initial conditions where the gravitational effects of the clock are negligible and therefore $C'\approx 0$ once $K_\varphi(x,0)$ is chosen, $E^\varphi$ is determined in order to satisfy the constraint and we recover a subset of the possible gauge choices of the complete theory. This includes all stationary gauges. Non stationary gauges can also be included. In fact, given $K_\varphi(x,t)$ when $C'\approx 0$, $E^\varphi(x,t)$ is determined by solving this equation and  the evolution equations are also satisfied in the $G \to 0$ limit.
When the Hamiltonian becomes bigger the initial $K_\varphi$ changes leading to non stationary solutions that differ more and more from the vacuum case when the energy grows. As expected, the introduction of the clock alters covariance by choosing the $t={\rm constant}$ surfaces. Covariance only becomes a weak coupling --small $G$-- approximation.

We will now discuss the quantization.

\section{Quantization}
The quantum treatment of the pure gravitational case is non-trivial. It requires the study of the spectrum of $H$ instead of solving the constraint, and to promote it to a self-adjoint operator. In particular, we will study the solutions that correspond to stationary gauge conditions and show how to recover previous results of the constrained system shown in \cite{usreview}. Let us recall that for a real clock, the momentum of the clock field $P_\phi$ is real and $C'$ must therefore be less than zero.  The quantization of $H$ requires, in order to be self-adjoint, to include all trajectories. For that we consider the spectrum of the $H$ operator, proportional to  $\sqrt{\vert C'\vert}$, and we will show that the solutions close to the fundamental state of the Hamiltonian system approximately reproduce the black hole geometry of the vacuum. The excited states differ more and more from the vacuum  due to the back-reaction of the clocks on the gravitational field.

Let us start with the kinematical quantum states we considered in our previous papers \cite{usreview}. Taking into account that the gauge fixing imposes that $E^x$ is a given function, and its conjugate $K_x$ is also determined by solving the diffeomorphism constraint, one needs to consider restricted spin networks that only involve $K_\varphi$ (see for instance  \cite{usreview}),
\begin{equation}
T_{g,\vec{\mu}}(K_\varphi) =
\prod_{v_j\in g}
\exp\left(i  {\mu_{j}} K_\varphi(v_j) \right).
\end{equation}

Where $g$ stands for the graph composed by the vertices $v_j$ that are located at $x_j$. The choice of $x_j$ corresponds to an identification of the radial coordinates with the quantized radius of the spheres of symmetry From now on we will omit $g$, since it remains fixed for all the states and use Dirac bra-ket notation with $T_{g,\vec{\mu}}(\vec{K}_\varphi) =\langle \vec{K}_\varphi|\vec{\mu}, M\rangle$, where $\vec{K}_\varphi$ corresponds to the sequence of $K_\varphi(v_i)$. Since the ADM mass is a Dirac observable, states are also labeled by its value $M$. 
Here we have a non singular Hamiltonian in a totally gauge fixed system and therefore the kinematical variables $E_\varphi$, $K_\varphi$  also are Dirac observables.
We start from the particular gauge fixing for $E^x(x_i)=\left( i\ell_{\rm Planck}\right)^2=k_i \ell_{\rm Planck}^2$, which corresponds to taking $x_i$ as a the radii of the spheres of symmetry and to choose a lattice with equally spaced radial coordinate $x_i=r_i=i\ell_{\rm Planck}$. Other time-independent choices of the radial coordinate are obviously possible. Time dependent choices might require further analysis. The index $i$ in principle can go from minus to plus infinity. In order to define the Hamiltonian operator we star by considering
\begin{equation}
    \hat{C}'_i=\frac{\hat{C}_{i+1}-\hat{C}_i}{\ell_{\rm Planck}},
\end{equation}
that in the $\mu_0$ Bohr compactification, with the polymerization parameter $\rho$ independent of $i$, takes the form:
\begin{equation}
    \hat{C}_i=-i \ell_{\rm Planck}
    \left(
    1+\frac{\sin^2\left(\rho \hat{K}_{\varphi,i}\right)}{\rho^2}
    -\frac{\left(2i+1\right)^2\ell_{\rm Planck}^2}{4 }\left(\hat{E}^\varphi_i\right)^{-2}+2GM
    \right),\label{29}
\end{equation}
with $\hat{K}_\varphi$ and $\hat{E}^\varphi$ represented by operators such that,
\begin{equation}
    \left(\frac{\sin\left(\rho \hat{K}_{\varphi,i}\right)}{\rho}\right)^2 \vert \vec{\mu},M\rangle =\frac{1}{4 \rho^2}
    \left(
    2 \vert \vec{\mu},M\rangle -
    \vert \ldots \mu_j+2 \rho\, \delta_{i,j} \dots,M\rangle-
    \vert \ldots \mu_j-2 \rho\, \delta_{i,j} \dots,M\rangle\right).
\end{equation}
and
\begin{equation}
    \left(\hat{E}^\varphi_i\right)^{-2} \vert \vec{\mu},M\rangle =\left( \left(\mu_i+\rho\right)^{2/3}-\left(\mu_i-\rho\right)^{2/3}\right)^6\left(\frac{3}{4\rho}\right)^6 \vert \vec{\mu},M\rangle,\label{31}
\end{equation}
where we have used the ``Thiemann trick''
which reproduces in the limit $\rho\to 0$ the quantity $1/\mu^2$. If we can determine the eigenstates $\psi_{\vec{\lambda}}$ of $\hat C$  that satisfy at each vertex,
\begin{equation}
    \hat{C}_i\psi_{\vec{\lambda}}\left(\vec{\mu}, M\right)=\lambda_i \psi_{\vec{\lambda}}\left(\vec{\mu},M\right)\label{32},
\end{equation}
then one can compute the action of the true Hamiltonian by noticing that  the term
${\hat h}_i$ containing the squared root acts as,
\begin{equation}
   {\hat h}_i\psi_{\vec{\lambda}}\left(\vec{\mu},M\right)= \left(\sqrt{\sqrt{E^x}\lvert\frac{C'}{\left(E^x\right)'}\rvert}\right)_i\psi_{\vec{\lambda}}\left(\vec{\mu},M\right)=\sqrt{\frac{i}{2i+1} \vert\lambda_{i+1}-\lambda_i\vert}\; \psi_{\vec{\lambda}}\left(\vec{\mu},M\right)\label{33}
\end{equation}
for $i>0$ and analogously for $i<0$.

Considering the operator associated with $H_{\rm true}$ given by equation (\ref{25}), its spectrum is continuous since in the representation $\vert \vec{\mu}, M\rangle$ the eigenvalues of $C_i$ constitute a continuous spectrum.

To recover the stationary solutions one has to seek normalizable states that minimize $H$. To this end, we can use a variational method.  For instance, we consider a linear combination of orthonormal states  $\Pi_i \psi_{b,a_n}^{x_i}(\mu_i)$  that contain the Gaussian  that approximate the polymerized version of the Schwarzschild metric in the Painlev\'e--Gullstrand gauge given by \cite{extensions},
\begin{equation}
    \psi_{b,a_n}^x(\mu)=\sum_n{a_n {H_n\left(\mu-\frac{x}{\ell_P}\right)}\left(\frac{2}{\pi b}\right)^{1/4}\exp\left(
    -\frac{1}{b}
    \left(\mu
    -\frac{x}{\ell_P}\right)^2\right)
    \exp{\left( i\mu \sqrt{\frac{2GM}{x}}\right)}},\;\label{34}
\end{equation}
(we drop the index $i$ for brevity)
with $\sqrt{{2GM}/{x}}$ the polymerized extrinsic curvature that takes its maximum value at the minimum value of $x_m$ that satisfies ${r_s}/{x_m} >{1}/{\rho}$ with $r_s=2GM$. That is the minimum $x_m>\rho 2GM$ which excludes the region that contains the singularity \cite{extensions}. ${H_n\left(\mu-\frac{x}{\ell_P}\right)}$ are Hermite polynomials.  
This family of states dependent on the parameters $b$ and $a_n$ includes a Gaussian approximation of the Schwarzschild black hole in Painlev\'e--Gullstrand coordinates for the case $a_0=1$ and $a_n=0$.

With this gauge fixed treatment all the dynamical variables that were not solved eliminating constraints are observables. So is the metric. Their associated operators can be written in terms of $E^\varphi$ and $K_\varphi$ and the Lagrange multipliers determined by the preservation of the gauge fixing constraints $N, N^x$.

In Painlev\'e--Gullstrand coordinates $E^\varphi=\sqrt{E^x}$ and classically, it only vanishes at the singularity. In order to apply the variational technique and define the square root appearing in $H_{\rm true}$ we need to determine the spectral decomposition of the operators ${\hat C}_i$ for each vertex $i$ given by equations (\ref{29}-\ref{32}). In order to avoid a lengthy calculation, in this first analysis we will consider the continuum approximation of the eigenvalue equation, which is valid for sufficiently small $\rho$'s if one excludes the surroundings of the singularity. Equation (\ref{32}), restricted to a given value of  $x=\sqrt{E^x}=i\ell_{\rm Planck}$ and $\rho/\mu\ll1$ is,
\begin{equation}
    \hat{C}_x \psi_x(\mu)=-x\left(\psi_x(\mu)-\frac{d^2}{d\mu^2} \psi_x(\mu)-
    \frac{\left(2x\right)^2(9\mu^2+22\rho^2)}{9 {\ell_{\rm Planck}}^2 \left(\mu^2+\rho^2\right)^2}\psi_x(\mu)\right)+\left(r_S-\lambda\right) \psi_x(\mu), \label{35}
\end{equation}
where we have included terms of order $\rho^2/\mu^4$ in the continuum approximation of $(E^\varphi)^{-2}$ 
We are interested in the normalizable solutions belonging in the continuous part of the spectrum. The simplest way of identifying them is to consider their asymptotic form for $\mu \to \infty$,  given by, 
\begin{equation}
    \hat{C}_x \psi_x(\mu)=-x\left(\psi_x (\mu)-\frac{d^2}{d\mu^2} \psi_x (\mu)
    \right)+\left(r_S-\lambda\right) \psi_x (\mu), \label{40}
\end{equation}
whose solutions for real $\lambda$ such that $\lambda < r_s-x$ are
\begin{equation}
    \psi_x^\lambda (\mu)=C_1 \sin\left(\mu\frac{\sqrt{-\lambda+r_S-x}}{\sqrt{x}}\right)
    +C_2 \cos\left(\mu\frac{\sqrt{-\lambda+r_S-x}}{\sqrt{x}}\right)\label{37}
\end{equation}
where $C_1$ and $C_2$ are constants.
For larger values of $\lambda$ the wavefunctions are not normalizable. The exact solutions of equation (\ref{35}) are
given by the Heun confluent functions, 
\begin{eqnarray}
    \psi_x^{\lambda}(\mu)&=&C_1 \left(\mu^2+\rho^2\right)^{\Lambda}   {\rm HeunC}
    \left(0,-\frac{1}{2},\Sigma,\rho^2\frac{\lambda-r_S+x}{4x},
    \Theta,-\frac{\mu^2}{\rho^2}\right)\nonumber\\
    &&
    +C_2 \left(\mu^2+\rho^2\right)^{\Lambda}{\rm HeunC}
    \left(0,\frac{1}{2},\Sigma,\rho^2\frac{\lambda-r_S+x}{4x},\Theta
    ,-\frac{\mu^2}{\rho^2}\right)\mu ,\\
    \Sigma&=&\frac{\sqrt{(52x^2+9{\ell_{\rm P}}^2)}}{6},\\
    \Theta&=&\frac{\left((18-9\rho^2)x+9\rho^2(rs-\lambda)\right){\ell_{\rm P}}^2+88x^3}{36x{\ell_{\rm P}}^2},\\
    \Lambda&=& \frac{\sqrt{(52x^2+9{\ell_{\rm P}}^2)}}{6\ell_{\rm P}}+\frac{1}{2}.
\end{eqnarray}
 These solutions form a continuous normalizable basis with elements $\psi_{\epsilon,x}^{\lambda}(\mu)$ that in the asymptotic region $\mu \to \infty$ take the form $\exp{\left(i\epsilon\mu\frac{\sqrt{\lambda-r_S+x}}{\sqrt{x}}\right)} $ with $\epsilon=\pm$. At each vertex $x=i\ell_{\rm Planck}$ we have, for each value of $\lambda$ two independent
basis elements $|\lambda, \epsilon>_{x}$. The basis elements satisfy  
\begin{equation}
{}_x<\lambda_1,\epsilon|\lambda_2.\epsilon'>_{x'}=\delta\left(\lambda_1-\lambda_2\right)\delta_{\epsilon,\epsilon'}\delta_{x,x'},
\end{equation}
and satisfy at the vertex located at $x$ the closure relation 
\begin{equation}
{\sum_\epsilon} \int_{-\infty}^{r_S-x} d \lambda\,\vert\lambda.\epsilon\rangle\langle\lambda,\epsilon\vert_x=I_x,
\end{equation}
where $I_x$ is the identity operator at this vertex.
Taking into account (\ref{11}) and (\ref{33}) and that the quantum Hamiltonian must be self-adjoint, we must consider the symmetrized product,
\begin{equation}
    \frac{1}{2}\left( \hat{E}^\varphi_i \hat{h}_i+\hat{h}_i\hat{E}^\varphi_i\right).
\end{equation}

We can then compute the expectation value of $\hat{H}$ for the multi-parametric family $\psi^x_{b,a_n}(\mu)$ given in (\ref{34}) for a fixed value of $x$,
\begin{equation}
    \langle \psi^x_{b,a_n}\vert \hat{H}\vert \psi^x_{b,a_n}\rangle = \sum_{i} L_i(b,a_n),
\end{equation}
where the sum in $i$ goes along all the vertices of the spin network, and,
\begin{eqnarray}
    L_i(b,a_n)&=&\int d\lambda_i\ldots d\lambda_N d\mu_1\ldots d\mu_N d\mu'_1\ldots d\mu'_N\nonumber\\
    &&\times
     \langle \psi^i_{b,a_n}\vert\mu'_1\cdots \mu'_N\rangle\left(\mu'_i+\mu_i\right)
    \left[\sqrt{f_i\left(\lambda_{i+1}-\lambda_i\right)}\Theta\left(\lambda_{i+1}-\lambda_i\right)
    +\sqrt{f_{i+1}\left(\lambda_i-\lambda_{i+1}\right)}\Theta\left(\lambda_i-\lambda_{i+1}\right)
    \right]\nonumber\\
    &&\times\langle \mu'_1\ldots \mu'_N\vert \lambda_1\ldots \lambda_N\rangle
    \langle\lambda_1\ldots\lambda_N\vert \mu_1\ldots \mu_N\rangle\langle \mu_1\ldots \mu_N\vert \psi^i_{b,a_n}\rangle>,
\end{eqnarray}
with 
\begin{equation}
    f_i=\frac{i}{\left(2i+1\right)\ell_{\rm Planck}}.
\end{equation}

Taking into account the closure relation for $\lambda$, $L_i(b)$ can be rewritten as,
\begin{eqnarray}
L_i(b,a_n)&=& \int d\lambda_i \lambda_{i+1} d\mu_i d\mu_{i+1} d\mu'_i d\mu'_{i+1}\nonumber\\
&&\times\langle\psi^i_{b,a_n}\vert \mu'_i \mu'_{i+1}\rangle\left(\mu'_i+\mu_i\right)
\left[\sqrt{f_i\left(\lambda_{i+1}-\lambda_i\right)}\Theta\left(\lambda_{i+1}-\lambda_i\right)
+\sqrt{f_{i+1}\left(\lambda_{i}-\lambda_{i+1}\right)}\Theta\left(\lambda_{i}-\lambda_{i+1}\right)
\right]\nonumber\\
&&\times
\langle \mu'_i\, \mu'_{i+1}\vert\lambda_i\, \lambda_{i+1}\rangle
\langle \lambda_i\, \lambda_{i+1}\vert\mu_i\, \mu_{i+1}\rangle\langle\mu_i\,\mu_{i+1}\vert\psi^i_{b,a_n}\rangle.
\end{eqnarray}

A numerical computation allows to show that with the exception of a small neighborhood of the bounce that occurs at $x$ of the order $\ell_{\rm Planck}$, the expectation value of the Hamiltonian is minimized for $a_n=0$  for non-vanishing $n$ and for a $b$ that grows linearly with $x$, and takes values $\langle{\hat H}\rangle = {K N \ell_{\rm Planck} }/{l_0^2}$ where $K$ is dimensionless and of order one, and $N$ the number of vertices of the spin network. Taking into account that for reasons of simplicity we have taken a non-local gauge fixing for the clock, the most natural choice for $l_0$ is the total extension of the lattice $x_M=N\ell_{\rm Planck}$. If this is the case then the integrated total true Hamiltonian is ---up to a dimensionless factor of order one--- proportional to the inverse of $x_M$. One therefore recovers, for the states that minimize the expectation value of the Hamiltonian, the Schwarzschild  black hole geometry. 
The previous calculation involves two drastic simplifications: a) We analyzed the continuum approximation of the original finite difference equation given in equations (\ref{29}-\ref{32}) and, b) instead of taking the exact solutions in terms of the Heun confluent functions we have taken the asymptotic wave functions given in (\ref{37}). The resulting expectation value is only an upper bound. However, even within this approximation the expectation value of $\dot K_\varphi$ for the wave-function that minimizes $<\hat H>$, vanishes. Therefore it is consistent to consider the choice of $K_\varphi$ as a stationary gauge choice as one has in Painlev\'e--Gullstrand coordinates. It should be noted that the quantum description is only consistent if both black holes ---corresponding to positive values of $x$ and $E^\varphi$--- and white holes ---corresponding to negative values--- are included. Only in that case the Hamiltonian operator is self-adjoint. Positive energy eigenstates describe the black hole and those of negative energy the white hole. The above analysis was performed in the black hole region where $x>0$.

\section{Conclusions}

We have shown that the treatment of the quantum self-gravitating scalar field here proposed, although it might be technically involved, is possible. The main technique required is the introduction of a basis of eigenstates of the radicand that appears in the Hamiltonian. We have therefore overcome obstacles that blocked the quantization of spherical models in loop quantum gravity including matter. We have identified the classical Hamiltonians that describe both the vacuum case as well as gravity coupled to a scalar field in spherical symmetry. We have presented possible techniques for their quantization based on spin networks, and show how to recover the Schwarzschild black hole geometry.

A rich set of options opens up, since the problem reduces to the treatment of a quantum Hamiltonian system with a discrete number of degrees of freedom. The resulting system is not unique as it depends on factor orderings that will have to be explored. Here we have carried out the vacuum case using a variational treatment to illustrate the new possibilities that are now available.

\section{Acknowledgements}
This work was supported in part by Grant NSF-PHY-1903799, NSF-PHY-2206557, funds of the
Hearne Institute for Theoretical Physics, CCT-LSU, Pedeciba, Fondo Clemente Estable
FCE 1 2019 1 155865.


\begin{thebibliography}{9}
\bibitem{usreview}R. Gambini, J. Olmedo, J. Pullin, 
[arXiv:2211.05621 [gr-qc]] to appear in "Handbook of Quantum Gravity", Cosimo Bambi, Leonardo Modesto, Ilya Shapiro (editors), Springer (2023)
\bibitem{choptuik} M.~W.~Choptuik,
Phys. Rev. Lett. \textbf{70}, 9-12 (1993)
doi:10.1103/PhysRevLett.70.9
\bibitem{gravityquantized}
M.~Domagala, K.~Giesel, W.~Kaminski and J.~Lewandowski,
Phys. Rev. D \textbf{82}, 104038 (2010)
doi:10.1103/PhysRevD.82.104038
[arXiv:1009.2445 [gr-qc]]; T. Thiemann 
[arXiv:astro-ph/0607380 [astro-ph]]; V.~Husain and T.~Pawlowski,
Phys. Rev. Lett. \textbf{108}, 141301 (2012)
doi:10.1103/PhysRevLett.108.141301
[arXiv:1108.1145 [gr-qc]];
K.~Giesel and T.~Thiemann,
Class. Quant. Grav. \textbf{32}, 135015 (2015)
doi:10.1088/0264-9381/32/13/135015
[arXiv:1206.3807 [gr-qc]].
\bibitem{florencia} F.~Ben\'\i{}tez, R.~Gambini, S.~L.~Liebling and J.~Pullin,
Phys. Rev. D \textbf{104}, no.2, 024008 (2021)
doi:10.1103/PhysRevD.104.024008
[arXiv:2106.00674 [gr-qc]].
\bibitem{usreview0} 
R.~Gambini, J.~Olmedo and J.~Pullin,
Class. Quant. Grav. \textbf{31}, 095009 (2014)
doi:10.1088/0264-9381/31/9/095009
[arXiv:1310.5996 [gr-qc]].
\bibitem{extensions}
R.~Gambini, J.~Olmedo and J.~Pullin,
Front. Astron. Space Sci. \textbf{8}, 74 (2021)
doi:10.3389/fspas.2021.647241
[arXiv:2012.14212 [gr-qc]].


\end{thebibliography}
\end{document}